\documentclass[10pt,a4paper]{article}
\usepackage[utf8]{inputenc}
\usepackage[english]{babel}
\usepackage{amssymb,graphicx}
\usepackage{amsmath,amsfonts}
\usepackage{amsthm}
\usepackage{framed}
\usepackage{fullpage}
\usepackage[makeroom]{cancel}
\usepackage[export]{adjustbox} 
\usepackage{feynmf}
\usepackage{microtype}
\usepackage{hyperref}
\usepackage{mathtools}

\def\be{\begin{eqnarray}}
\def\ee{\end{eqnarray}}
\def\nn{\nonumber}

\def\p{\partial}

\newcommand{\beq}{\begin{equation}}
\newcommand{\eeq}{\end{equation}}
\newcommand{\beqa}{\begin{eqnarray}}
\newcommand{\eeqa}{\end{eqnarray}}

\def\NS{\mathfrak{P}}
\def\Z{x}

\usepackage[cmtip,all]{xy}

\newcommand{\longsquiggly}{\xymatrix{{}\ar@{~>}[r]&{}}}

\begin{document}

\title{\vspace{1.5cm}\bf
Elliptic triad
}

\author{
A. Mironov$^{b,c,d,}$\footnote{mironov@lpi.ru,mironov@itep.ru},
A. Morozov$^{a,c,d,}$\footnote{morozov@itep.ru},
A. Popolitov$^{a,c,d,}$\footnote{popolit@gmail.com},
Z. Zakirova$^{e,d}$\footnote{zolya\_zakirova@mail.ru}
}

\date{ }

\maketitle

\vspace{-6cm}

\begin{center}
\hfill MIPT/TH-24/24\\
\hfill  FIAN/TD-15/24\\
\hfill ITEP/TH-39/24 \\
\hfill IITP/TH-33/24
\end{center}

\vspace{3cm}

\begin{center}
$^a$ {\small {\it MIPT, Dolgoprudny, 141701, Russia}}\\
$^b$ {\small {\it Lebedev Physics Institute, Moscow 119991, Russia}}\\
$^c$ {\small {\it NRC ``Kurchatov Institute", 123182, Moscow, Russia}}\\
$^d$ {\small {\it Institute for Information Transmission Problems, Moscow 127994, Russia}}\\
$^e$ {\small {\it Kazan State Power Engineering University, Kazan, Russia}}
\end{center}

\vspace{.1cm}

\begin{abstract}
The triad refers to embedding the Macdonald polynomials into the Noumi-Shiraishi functions
and their reduction to solutions of simple linear equations at particular values of $t$.
It provides an alternative definition of Macdonald theory.
We discuss lifting the triad to an elliptic generalization of the Noumi-Shiraishi functions.
The central unknown ingredient is linear equations, for which we discuss
various possible approaches, including immediate elliptic deformation of periodicity conditions,
(elliptic) Ding-Iohara-Miki algebra operators, and elliptic Kostka coefficients.
\end{abstract}

\bigskip

\newcommand\smallpar[1]{
  \noindent $\bullet$ \textbf{#1}
}

\section{Introduction}

The Macdonald polynomials are a puzzling discovery \cite{Mac}, which motivates various developments
in mathematical physics for decades.
They are close relatives of the Schur polynomials, which are characters of representations of $sl_N$ group,
but do not have a simple group theory interpretation.
Only recently it became clear that the Macdonald polynomials belong to representation theory of a rather
complicated Ding-Iohara-Miki (DIM) algebra \cite{DIM}, though details still remain under investigation.
At the same time, the Macdonald polynomials long played a role in the study of physical systems,
from integrable systems, where they are eigenfunctions of the Ruijsenaars Hamiltonians
(now known to be among elements of the DIM algebra)
to knot (refined Chern-Simons) theory, where they provide {\it hyper}polynomials.
Despite Macdonald theory is still far from being completed and fully understood,
attempts are made for its further generalizations, the most successful is to
Noumi-Shiraishi (NS) functions, which are no longer polynomials, and are no longer labeled
by Young diagrams (partitions).
On another side, Macdonald functions can be considered as an analytical continuation in the parameter $t$
of conceptually simpler objects: solutions to linear systems, which are determinants
of some simply-looking but peculiar functions.
These two developments are in fact interrelated, and recently we suggested \cite{MMP3} to name the
whole entity a {\it triad},
emphasizing its three sources and three component parts:

\begin{itemize}
\item The NS function. It is defined to be a power series in $x_i=q^{z_i}$ depending on $N$ complex parameters $y_i=q^{\lambda_i}$ and on two parameters $q$ and $t$. It is an eigenfunction of the Macdonald-Ruijsenaars operator \cite{NS}. It is invariant under the Poincar\'e duality $t\to q/t$ and (up to some simple factors)  under the Ruijsenaars duality $x_i\leftrightarrow y_i$. The NS function can be also lifted to the Shiraishi function \cite{Shi} that solves the non-stationary elliptic Ruijsenaars equation.
\item The Macdonald polynomial \cite{Mac}. It is a symmetric polynomial in $x_i$ depending on two parameters $q$ and $t$ and labeled by the Young diagram (partition) $\mu=\{\mu_i\}$.  It is still an eigenfunction of the Macdonald-Ruijsenaars operator
\item The Baker-Akhiezer (BA) function. It can be defined as a (quasi)polynomial of a given power (see (\ref{gsBA2})) in $x_i$ depending on $N$ complex parameters $y_i$ and on $q$ and $t=q^{-m}$ ($m\in\mathbb{Z}_{>0}$), which satisfies periodicity conditions (see (\ref{symm1})), and these conditions are enough to fix the BA function unambiguously up to a constant normalization factor \cite{Cha}. A corollary of this construction is that this BA function is an eigenfunction of the Macdonald-Ruijsenaars operator \cite{Cha}. The periodicity conditions can be rewritten as a system of linear equations that we mentioned above. Moreover, the BA function can be presented by a kind of Rodrigues formula (see (\ref{prod})).
\end{itemize}
Both the Macdonald polynomial and the BA function are obtained from the NS function by reductions: the Macdonald polynomial, by specifying $y_i=q^{\mu_i}t^{N-i}$ with $\mu_i$ forming a partition, and the BA function, by specifying $t=q^{-m}$ ($m\in\mathbb{Z}_{>0}$).

In the present letter, we make one more step emphasizing the significance of triad:
all the three ingredients allow elliptic deformations\footnote{By the words ``elliptical deformation"
we mean a deformation of the coefficients only so that the polynomials remain to be polynomials.} in three different directions, and all three continue to be related.
That is,

\begin{itemize}
\item The elliptic NS function \cite{FOS,AKMM2,AKMM3,CSW,Zen}. It depends on an additional  parameter (elliptic nome) $w$. It is an eigenfunction of the degenerate elliptic-trigonometric Koroteev-Shakirov Hamiltonians \cite{MMZ}. The elliptic NS function can be also extended to the elliptic lift of the Shiraishi (ELS) function \cite{AKMM2}, which, at a particular (stationary) limit, is an eigenfunction of the full Koroteev-Shakirov Hamiltonians \cite{MMZ}, and celebrates a series of dualities \cite{AKMM2}.
\item The elliptic Macdonald polynomial \cite{FOS,AKMM2,AKMM3}. It is a symmetric polynomial in $x_i$ depending on parameters $q$, $t$ and $w$ and labeled by the Young diagram (partition) $\mu=\{\mu_i\}$. The elliptic Macdonald polynomial celebrates a series of important properties, in particular, polynomials orthogonal to the elliptic Macdonald polynomials with respect to the Schur scalar product are eigenfunctions of the Hamiltonians dual to the elliptic Ruijsenaars Hamiltonians \cite{MMZ}.
\item The Baker-Akhiezer (BA) function. It can be defined as a (quasi)polynomial of a given power (see (\ref{gsBA2})) in $x_i$ depending on $N$ complex parameters $y_i$ and on $q$, $t=q^{-m}$ ($m\in\mathbb{Z}_{>0}$) and $w$, and it is still an eigenfunction of the degenerate elliptic-trigonometric Koroteev-Shakirov Hamiltonians. However, counterparts of both the periodicity conditions and of a Rodrigues type formula have not been found yet as we explain in sec.3.
\end{itemize}

As we explain in sec.2.2, the elliptic deformation of the triad is very restrictive because of a series of non-trivial identities that have to be satisfied in order to guarantee reduction to symmetric polynomials.

Besides conceptual importance,
the elliptic triad attracts new attention to ELS functions,
a very important and challenging chapter of modern theory.

The plan of our letter is as follows. In section 2, we describe all three ingredients of the elliptic triad and interrelations between them. In section 3, we discuss the problems of obtaining periodicity conditions and of a Rodrigues type formula for the elliptic BA functions. In section 4, we introduce the orthogonality relations for the elliptic Macdonald polynomials and describe properties of their duals under the Schur scalar product. We also emphasize problems of constructing orthogonality relations for the BA functions. Section 5 contains some concluding remarks.

\paragraph{Notation.}
The Pochhammer symbols are defined to be
\be
(z;q)_\infty:=\prod_{n=0}^\infty (1-zq^n),\ \ \ \ \ \ \ (z;q,w)_\infty:=\prod_{n,m=0}^\infty (1-zq^nw^m)
\ee
We need the odd $\theta$-function that we define as\footnote{It differs by a factor from the standard odd $\theta$-function \cite{BE}:
\be
\theta_1(u;\tau)={iw^{1/8}\cdot (w;w)_\infty\over \sqrt{z}}\cdot\theta_{w}(z)\Big|_{w=e^{2\pi i\tau},z=e^{2\pi i u}}
\ee}
\be
\theta_w(z):=(z;w)_\infty(w/z;w)_\infty={1\over (w;w)_\infty}\sum_{n\in\mathbb{Z}}(-1)^n
w^{n(n-1)\over 2}z^n
\ee
and the even $\theta$-function (which is $\theta_3$)
\be
\theta_w^{(e)}(z):=\sum_{n\in\mathbb{Z}}
w^{n^2\over 2}z^n
\ee

The elliptic $\Gamma$-function is defined to be
\be
\Gamma(z;q,w):={(qw/z;w,q)_\infty\over (z;w,q)_\infty}=\exp\left[\sum_m{z^m-(wq/z)^m\over (1-q^m)(1-w^m)m}\right]
\ee
The elliptic Pochhammer symbol is defined as
\beq\label{Theta}
\Theta(z;q,w)_n: ={\Gamma(q^nz;q,w)\over\Gamma(z;q,w)}
\eeq

\section{Elliptic triad}

\subsection{A basic object: elliptic NS function}

The elliptic NS function \cite{AKMM2,AKMM3,FOS} is, up to a simple pre-factor, a formal power series of $2N$ variables $x_i=q^{z_i}$ and $y_i=q^{\lambda_i}$, $i=1,\ldots,N$ generically neither symmetric, nor polynomial in $q^{z_i}$. However, it can acquire one, or both, of these properties at peculiar specializations. This power series is defined to be
\be\label{NS}
\NS_{q,t,w}(\vec z,\vec \lambda)= q^{\vec z\cdot\vec\lambda}\cdot t^{-\vec z\cdot\vec\rho}\cdot
\left(\sum_{k_{ij}}\psi(\vec\lambda,k_{ij};q,t,w)\prod_{1\le i<j\le N}q^{k_{ij}(z_j-z_i)}\right)
\ee
where the sum goes over all non-negative integer $k_{ij}$ with $i< j$,
and $\vec\rho$ is the Weyl vector, i.e. $\vec\rho\cdot\vec z={1\over 2}\sum_{i=1}^N(N-2i+1)z_i$. The coefficients $\psi(\vec\lambda,k_{ij};q,t)$ are
\be
\psi(\vec\lambda,k_{ij};q,t,w):&=&\prod_{n=2}^N\prod_{1\le i<n}\prod_{s=0}^{k_{in}-1}{\theta_w\Big(q^{s+1-k_{in}}t^{-1}\Big)
\over \theta_w\Big(q^{s-k_{in}}\Big)}
{\theta_w\Big(tq^{s+\lambda_n-\lambda_i+\sum_{a>n}(k_{ia}-k_{na})}\Big)
\over \theta_w\Big(q^{s+1+\lambda_n-\lambda_i+\sum_{a>n}(k_{ia}-k_{na})}\Big)}
\times\\
&\times&
\prod_{n=2}^N\prod_{1\le i<j< n}\prod_{s=0}^{k_{in}-1}{\theta_w\Big(tq^{s+\lambda_j-\lambda_i+\sum_{a>n}(k_{ia}-k_{ja})}\Big)
\theta_w\Big(t^{-1}q^{s+1+\lambda_j-\lambda_i-k_{jn}+\sum_{a>n}(k_{ia}-k_{ja})}\Big)
\over \theta_w\Big(q^{s+1+\lambda_j-\lambda_i+\sum_{a>n}(k_{ia}-k_{ja})}\Big)
\theta_w\Big(q^{s+\lambda_j-\lambda_i-k_{jn}+\sum_{a>n}(k_{ia}-k_{ja})}\Big)}\nn
\label{c}
\ee
The elliptic NS function is an eigenfunction of the degenerate elliptic-trigonometric Koroteev-Shakirov Hamiltonians \cite{MMZ,MMdell}. They are better to present as a single operator depending on two parameters and realized in terms of an operator $\hat{\cal O}(u)$ that admits a determinant form \cite{ZG},
\be\label{O}
\hat{\cal O}(u):={1\over\Delta(q^{\vec z})}\det_{1\le i,j\le N}
\Big(q^{(N-j)z_i}\theta_w(ut^{N-j}e^{\p_{z_i}}\Big)
\ee
where $\Delta(q^{\vec z})=\prod_{i<j}(x_i-x_j)$ is the Vandermonde determinant, so that
\be\label{MR}
\hat H^{etKS}(u,v)=\hat{\cal O}(u)\Big(\hat{\cal O}(v)\Big)^{-1}\\
\hat H^{etKS}(u,v)\NS_{q,t,w}(\vec z,\vec \lambda)=\prod_{i=1}^N
  \frac{\theta_{\omega}(v q^{\lambda_i}t^{N-i})}{\theta_{\omega}(u
    q^{\lambda_i}t^{N-i})}\cdot\NS_{q,t,w}(\vec z,\vec \lambda)
\ee
These operators can be understood both in terms of DIM algebra elements (see \cite{FOS} and \cite{Zen} for different realizations) and in terms of conjugate Hamiltonians of the elliptic DIM algebra \cite{MMZ}.

\paragraph{An example of $N=2$.}

In this case, there is just one non-zero $k_{12}=k$, and, using the notation $z:=z_1-z_2$, $\lambda:=\lambda_1-\lambda_2$, one obtains
\be\label{NS2}
\NS_{q,t,w}(z_1,z_2,\lambda_1,\lambda_2)=q^{\lambda_1z_1+\lambda_2 z_2}\cdot t^{-z\over 2}\cdot
\sum_{k=0}\psi(\lambda_1-\lambda_2,k;q,t,w)q^{k(z_2-z_1)}
\ee
with
\be
\psi(\lambda,k;q,t,w)=\prod_{s=0}^{k-1}{\theta_w\Big(q^{s+1-k}t^{-1}\Big)
\over \theta_w\Big(q^{s-k}\Big)}{\theta_w\Big(tq^{s-\lambda}\Big)
\over \theta_w\Big(q^{s+1-\lambda}\Big)}=\prod_{s=1}^{k}{\theta_w\Big(q^{s-1}t\Big)
\over \theta_w\Big(q^{s}\Big)}{\theta_w\Big(t^{-1}q^{\lambda-s+1}\Big)
\over \theta_w\Big(q^{\lambda-s}\Big)}
\label{c2}
\ee

In this case, there is only one non-trivial Hamiltonian, it can be constructed instead of (\ref{MR}) from two operators \cite{KS,AKMM1,MMdell},
\be\label{H2e}
\hat{\mathfrak{O}}_0&=&\sum_{n\in\mathbb{Z}}w^{n^2\over 2}\sinh{z_1-z_2-2mn\over 2}
e^{n\p_1-n\p_2}\sim\nn\\
&\sim&\sqrt{x_1\over x_2}\theta_w^{(e)}\Big(q^{x_1\p_1-x_2\p_2-m}\Big)
-\sqrt{x_2\over x_1}\theta_w^{(e)}\Big(q^{x_1\p_1-x_2\p_2+m}\Big)
\\
\hat{\mathfrak{O}}_1&=&\sum_{n\in\mathbb{Z}}w^{n(n-1)\over 2}\sinh{z_1-z_2-m(2n-1)\over 2}
e^{n\p_1-(n-1)\p_2}\sim\nn\\
&\sim&\sqrt{q^mx_1\over x_2}
\theta_w\Big(q^{x_1\p_1-x_2\p_2-m}\Big)q^{x_2\p_2}
-\sqrt{x_2\over q^mx_1}\theta_w\Big(q^{x_1\p_1-x_2\p_2+m}\Big)q^{x_2\p_2}\nn
\ee
with $\sinh(x):={q^x-q^{-x}\over 2}$, and is equal to
\be
\hat H=\hat{\mathfrak{O}}_0^{-1}\hat{\mathfrak{O}}_1
\ee

\subsection{Reduction to elliptic Macdonald polynomials}

As was proved in \cite{MMZ}, the polynomial eigenfunctions of the Hamiltonian (\ref{MR}) are the elliptic Macdonald polynomials \cite{AKMM2}. Choosing other boundary conditions, one can obtain more solutions. The elliptic NS functions provides an example of solutions that are power series instead of polynomials, they are not symmetric functions, in contrast with the elliptic Macdonald polynomials, but instead they are parameterized by $N$ arbitrary complex parameters, while, in the Macdonald case, there is an integrality requirement.

However, the power series (\ref{NS}) can be made a symmetric polynomial in $q^{z_j}$ by choosing $\vec\lambda=\vec\mu-\vec\rho\log_q t$, where $\vec\mu$ has all non-negative integer components $\mu_j$, $\mu_1\ge\mu_2\ge\ldots\ge\mu_N\ge0$ at $j=1,\ldots,N$. This symmetric polynomial is nothing but the elliptic Macdonald polynomial \cite{NS}:
\be\label{NSM}
\NS_{q,t,w}(\vec z,\vec \mu-\vec\rho\log_q t)= P_\mu(\{q^{z_i}\};q,t,w)
\ee
Notice the peculiar $t$-dependent shift: it turns out to be crucial
in non-commutativity of triad specializations, see sec.\ref{sec:two-non-perm-reductions}.

Note that the fact that the polynomial $P_\mu(\{x_i\};q,t,w)$ is symmetric and $N$-independent, i.e. admits a reformulation in terms of power sums $p_k:=\sum_{i=1}^Nq^{kz_i}$ is a non-trivial corollary of a series of elliptic identities \cite{AKMM2}. For instance, in the simplest non-trivial case of $N=4$, the symmetricity of $P_{[3,1]}(\{x_i\};q,t,w)$ requires that
\be\label{1st}
\zeta_2(1)^2-\zeta_2(1)\zeta_2(qt)-\zeta_2(1)\zeta_2(q)+\zeta_2(t)\zeta_2(qt)=0\\
\zeta_k:={\theta_w(q^kz)\theta_w(tz)\over\theta_w(q^{k-1}tz)\theta_w(qz)}
\ee
while that of $P_{[4,1]}(\{x_i\};q,t,w)$, (\ref{1st}) and also
\be
\zeta_2(1)\zeta_3(1)-\zeta_2(q^2t)\zeta_3(1)-\zeta_4(1)+\zeta_4(t)=0\label{2nd}\\
\zeta_3(1)-\zeta_2(1)\zeta_2(q^2t)-\zeta_3(q)+\zeta_3(qt)=0
\ee
etc.
These identities are, indeed, true for the theta-functions. Moreover, if one considers instead of the $\theta_w(x)$ an arbitrary function $\eta(x)$ with the property $\eta(x^{-1})=-\eta(x)/x$ (which is a necessary consistency condition), which can be expanded into the Taylor series at $x=1$, one realizes that {\bf all} solutions to these identities are parameterized by two parameters, and one of these parameters is just the normalization factor. Hence, one obtains that only $\eta(x)\sim\theta_w(x)$ solves these identities so that the elliptic triad exists only in this case!

\paragraph{An example of $N=2$.}

In this case, using the notation $x_{1,2}:=q^{z_{1,2}}$, $\mu:=\mu_1-\mu_2$, one obtains from (\ref{NS2}), (\ref{c2}) and (\ref{NSM}) (see \cite[Eq.(5.1)]{KS}):
\be\label{P2}
P_{[\mu_1,\mu_2]}(\{x_1,x_2\};q,t,w)= x_1^{\mu_1}x_2^{\mu_2}\cdot
\sum_{k=0}\left({x_2\over x_1}\right)^{k}\prod_{s=1}^{k}{\theta_w\Big(q^{s-1}t\Big)
\over \theta_w\Big(q^{s}\Big)}{\theta_w\Big(q^{\mu-s+1}\Big)
\over \theta_w\Big(tq^{\mu-s}\Big)}
\ee

\subsection{Reduction to elliptic BA functions}

Instead of getting symmetric polynomials with integrality requirements for $\vec\lambda$, one can choose $t=q^{-m}$, which also gives rise to a (quasi)polynomial\footnote{We use the term quasipolynomial because of a generally non-polynomial trivial pre-factor $q^{\vec\lambda\cdot\vec z}$.} of $q^{z_i}$ though non-symmetric, but instead $\vec\lambda$ can be kept to be $N$ arbitrary complex parameters. Hence, on one hand, it does not admit a reformulation in terms of power sums $p_k:=\sum_{i=1}^Nq^{kz_i}$, but, on the other hand, this (quasi)polynomial is nothing but the elliptic BA function $\Psi_m(\vec z,\vec\lambda;q,w)$ \cite{Cha}:
\be\label{NBA}
\Psi_m(\vec z,\vec\lambda;q,w)={\cal N}_\lambda\cdot\NS_{q,q^{-m},w}(\vec z,\vec \lambda)
\ee
where ${\cal N}_\lambda$ is a normalization factor. Unfortunately, in variance with the non-elliptic case, one can not make the BA function a symmetric function in $\vec z$ and $\vec\lambda$ by choosing this normalization factor. This was possible for the ordinary BA function because it was an eigenfunction of the self-dual Macdonald-Ruijsenaars Hamiltonian, and the Hamiltonian (\ref{MR}) is not self-dual, and the elliptic BA function is its eigenfunction.

\paragraph{Elliptic Macdonald polynomials from the elliptic BA functions.} One can make the elliptic Macdonald polynomial from the BA function at non-negative integer components of $\vec\lambda+m\vec\rho$ ordered in non-increasing order (and this is Chalykh's original relation between the two non-elliptic quantities, see \cite[Thm.5.11]{Cha}):
\be\label{MPsi}
P_{\vec\lambda+m\vec\rho}(\{q^{z_i}\};q,q^{-m},w)=
{\cal N}_\lambda^{-1}\cdot\sum_{w\in W}\Psi_m(w\vec z,\vec\lambda;q,w)
\ee
Here $w$ is an element of the Weyl group of $A_{N-1}$.

Note also that, in the elliptic case, there is no the Poincar\'e symmetry $t\to q/t$ of the elliptic NS function (see \cite[Theorem 6.5]{NS} for the non-elliptic case), and, hence,
one can not similarly construct \cite{Cha} the Macdonald polynomial at $t=q^{m+1}$.

\section{Properties of the elliptic BA function}

\subsection{On two non-permutable reductions from the elliptic NS function\label{sec:two-non-perm-reductions}}

Let us stress that, making the reduction (\ref{NSM}) at discrete values of $\vec\lambda$, one obtains a symmetric polynomial, while making the reduction (\ref{NBA}) at $t=q^{-m}$, one obtains a non-symmetric polynomial that does not become symmetric after further restricting $\vec\lambda$ according to (\ref{NSM}). In other words, the two reductions are not permutable.

Let us demonstrate how this works in the simplest case of $N=2$ (\ref{NS2}). In this case, the power series
\be
x_1^{\mu_1}x_2^{\mu_2}\cdot
\sum_{k=0}\left({x_2\over x_1}\right)^{k}\prod_{s=1}^{k}{\theta_w\Big(q^{s-1}t\Big)
\over \theta_w\Big(q^{s}\Big)}{\theta_w\Big(q^{\mu-s+1}\Big)
\over \theta_w\Big(tq^{\mu-s}\Big)}
\ee
becomes the elliptic Macdonald polynomial $P_{[\mu_1,\mu_2]}$ at integer $\mu_{1,2}$, $\mu=\mu_1-\mu_2\ge 0$ at generic values of the parameter $t$. The sum in this case is limited by $k=\mu$ because of vanishing the second multiplier in the nominator at larger $k$. Then, choosing in this elliptic Macdonald polynomial $t=q^{-m}$ and $\mu>2m$, we leave in the sum the terms with $0\le k\le m$ and with $\mu-m\le k\le \mu$.

However, if one chooses $t=q^{-m}$ from the very beginning keeping $\mu$ generic, the expression becomes
\be\label{2}
x_1^{\mu_1}x_2^{\mu_2}\cdot
\sum_{k=0}\left({x_2\over x_1}\right)^{k}\prod_{s=1}^{k}{\theta_w\Big(q^{s-1-m}\Big)
\over \theta_w\Big(q^{s}\Big)}{\theta_w\Big(q^{\mu-s+1}\Big)
\over \theta_w\Big(q^{\mu-s-m}\Big)}
\ee
and only the terms with $0\le k\le m$ contribute because of vanishing the first multiplier in the nominator at larger $k$. This gives rise to the elliptic BA function, which is defined at arbitrary $\mu$.

The point is that, at positive integer $\mu$, there are cancellations between the first multiplier in the nominator and the second multiplier in the denominator at $\mu-m\le k\le \mu$. Hence, the elliptic Macdonald polynomial defined only at such values of $\mu$, contains these terms, and the elliptic BA function defined at arbitrary $\mu$ does not.

Thus, we demonstrated that making a reduction of the elliptic NS function at $t=q^{-m}$ and arbitrary $\lambda$ restricts the obtained polynomial to a Weyl chamber, while making a reduction at $\vec\mu=\vec\lambda-m\vec\rho$ forming a partition gives rise to a complete symmetric polynomial.

Note that we fix a concrete Weyl chamber choosing in the elliptic NS function the ordering $i<j$ in (\ref{NS}) and the pre-factor $q^{\vec z\cdot\vec\lambda}$. This fixes the procedure (\ref{NSM}) of reduction to the elliptic Macdonald polynomial, when the symmetry between different $x_i$ is restored. At the same time, this choice in the elliptic NS function also fixes a concrete Weyl chamber associated with the elliptic BA function. One definitely can make any other choice.

\subsection{Basic definition of the BA function}

In the non-elliptic case, one defines the BA function from two requirements: one requires the BA function to be of the form
\be\label{gsBA}
\Psi_m(\vec z,\vec\lambda)=q^{(\vec\lambda+m\vec\rho) \vec z}\ \sum_{k_{ij}=0}^mq^{-\sum_{i>j}k_{ij}(z_i-z_j)}\psi_{m,\vec\lambda,k}
\ee
and satisfy the periodicity conditions
\be\label{symm1}
\Psi_m(z_k+j,\lambda)=\Psi_m(z_l+j,\lambda)\ \ \ \ \  \forall k,l\ \ \hbox{and}\ \ 1\le j\le m\ \ \ \ \ \hbox{at}\ \ q^{z_k}=q^{z_l}
\ee
These requirements unambiguously fix the BA function (up to a common normalization factor).
Technically, one inserts the anzatz (\ref{gsBA}) into (\ref{symm1}) and obtains a set of linear equations giving the coefficients $\psi_{m,\vec\lambda,k}$. As we explain below, {\bf the periodicity conditions leads} to all essential properties of the BA functions: they are {\bf eigenfunctions of the Macdonald-Ruijsenaars operator}, they can be obtained by a kind of {\bf Rodrigues formulas}, etc. Unfortunately, a counterpart of the periodicity properties and of the linear equations in the elliptic case is not clear, and the elliptic BA functions are given only by explicit formulas. It results into problems with obtaining various properties of the elliptic BA functions, as we discuss below.

\subsection{Linear equations}

In the simplest example of $N=2$, the BA function
\be\label{N=2}
\Psi_m(x_1,x_2,\vec\lambda)=x_1^{\lambda_1}x_2^{\lambda_2}\ \sum_{k=0}^m\left({x_2\over x_1}\right)^{k-{m\over 2}}\psi_{m,\vec\lambda,k}
\ee
can be effectively reduced \cite{MMP1} to
\be\label{Psixi}
\overline{\Psi}_m(\xi,\lambda)= \sum_{k=0}^m\xi^{{m+\lambda\over 2}-k}\psi_{m,\vec\lambda,k},
\ \ \ \ \ \ \xi:={x_1\over x_2}
\ee
and (\ref{symm1}) can be written in the form
\be\label{symxi}
\overline{\Psi}_m(q^j,\lambda)=\overline{\Psi}_m(q^{-j},\lambda),\ \ \ \ \ \ \ j=1,\ldots,m
\ee
Then, one can rewrite this linear equations in the form
\be
\sum_{k=0}^m\left(\xi_j^{k-{\lambda+m\over 2}}-\xi_j^{{\lambda+m\over 2}-k}\right)\psi_{m,\lambda,k}=0\ \ \ \ \  1\le j\le m
\ee
with $\xi_j=q^j$, $j=1,\ldots,m$.
This form admits an extension to the elliptic case at $m=1$:
\be\label{leq}
\sum_{k=0}^m\theta_w(\xi_j^{2k-\lambda- m})\xi_j^{{\lambda+m\over 2}-k}\psi^{ell}_{m,\lambda,k}=0\ \ \ \ \  1\le j\le m
\ee
Indeed, one needs to obtain
\be\label{ellcf}
\psi^{ell}_{m,\vec\lambda,k}=\prod_{j=1}^{k}{\theta_w(q^{\lambda+m-j+1})\theta_w(q^{j-1-m})
\over\theta_w(q^{j})\theta_w(q^{\lambda-j})}
\ee
i.e.
\be
\psi^{ell}_{m,\vec\lambda,0}=1\nn\\
\psi^{ell}_{m,\vec\lambda,1}={\theta_w(q^{\lambda+m})\theta_w(q^{-m})
\over\theta_w(q)\theta_w(q^{\lambda-1})}
\ee
and, at $m=1$, condition (\ref{leq}) becomes
\be
\theta_w(q^{-\lambda-1})q^{\lambda+1\over 2}+\theta_w(q^{1-\lambda})q^{\lambda-1\over 2}
{\theta_w(q^{\lambda+1})\theta_w(q^{-1})
\over\theta_w(q)\theta_w(q^{\lambda-1})}=0
\ee
where we used that $\theta_w(x^{-1})=-{\theta_w(x)\over x}$.

Unfortunately, extension to higher $m$ is ambiguous (there are many different equations with the same solution) and an additional criterium is needed to make a selection.

\subsection{Deriving linear equations}

Let us see why the equation that is satisfied by the BA function is consistent with the periodicity conditions (\ref{symxi}) at $N=2$ case. The key observation is that the Macdonald-Ruijsenaars operator \cite{Rui},
\be
\hat D_\Z:=\sum_{i=1}^N\prod_{j\ne i}{\Z_j-q^{-m}\Z_i\over \Z_j-\Z_i}e^{\p_{z_i}}
\ee
preserves the periodicity property of functions (\ref{symxi}). Indeed, it acts on the function of $f(\xi)$ in accordance with
\be
\hat D_\xi f(\xi)={1-q^{-m}\xi\over 1-\xi}f(q\xi)+{1-q^{-m}\xi^{-1}\over 1-\xi^{-1}}f(q^{-1}\xi)
\ee
and, hence,
\be\label{1}
\hat D_\xi f(q^j)={1-q^{-m+j}\over 1-q^j}f(q^{j+1})+{1-q^{-m-j}\over 1-q^{-j}}f(q^{j-1})\\
\hat D_\xi f(q^{-j})={1-q^{-m-j}\over 1-q^{-j}}f(q^{1-j})+{1-q^{-m+j}\over 1-q^{j}}f(q^{-j-1})
\label{2}
\ee
Assume now that $f(q^j)=f(q^{-j})$ at $j=1,\ldots,m$. Then, one observes that, at $j=1,\ldots,m-1$:
\be
\hat D_\xi f(q^{-j})={1-q^{-m-j}\over 1-q^{-j}}f(q^{1-j})+{1-q^{-m+j}\over 1-q^{j}}f(q^{-j-1})=
{1-q^{-m-j}\over 1-q^{-j}}f(q^{j-1})+{1-q^{-m+j}\over 1-q^{j}}f(q^{j+1})=
\hat D_\xi f(q^j)
\ee
and, at $j=m$,
\be
\hat D_\xi f(q^{-m})={1-q^{-2m}\over 1-q^{-m}}f(q^{1-m})={1-q^{-2m}\over 1-q^{-m}}f(q^{m-1})=
\hat D_\xi f(q^m)
\ee
In order to generalize this trick to the elliptic case, one has to find a counterpart of the periodicity property that is preserved under the action of the operator $\hat{\cal O}(u)$ (\ref{O}), which, at $N=2$, reads
\be
\hat{\cal O}(u)={1\over x_1-x_2}\left[x_1\theta_w\Big(uq^{-m}q^{x_1\p_{x_1}}\Big)
\theta_w\Big(uq^{x_2\p_{x_2}}\Big)-x_2\theta_w\Big(uq^{-m}q^{x_2\p_{x_2}}\Big)
\theta_w\Big(uq^{x_1\p_{x_1}}\Big)\right]
\ee
Up to a trivial factor, this Hamiltonian acts on the function $f(x_1,x_2)$ as
\be
\hat{\cal O}(u)f(x_1,x_2)=\sum_{n,\bar n\in\mathbb{Z}}(-u)^{n+\bar n}
w^{n(n-1)+\bar n(\bar n-1)\over 2}{x_1q^{-mn}-x_2q^{-m\bar n}\over x_1-x_2}
f(q^nx_1,q^{\bar n}x_2)
\ee
and, on the function $f(x_1/x_2)=f(\xi)$ as
\be\label{Opc}
\hat{\cal O}(u)f(\xi)=\sum_{n,\bar n\in\mathbb{Z}}(-u)^{n+\bar n}
w^{n(n-1)+\bar n(\bar n-1)\over 2}{\xi q^{-mn}-q^{-m\bar n}\over \xi-1}
f(q^{n-\bar n}\xi)
\ee
In the limit of $w\to 0$, only the four terms with $n,\bar n=0,1$ contributes, and one obtains
\be
\hat{\cal O}(u)\Big|_{w=0}f(\xi)=(1+u^2q^{-m})f(\xi)-u\hat D_\xi f(\xi)
\ee
Note that the periodicity condition should admit a non-trivial solution. For instance, it is clear that (\ref{Opc}) does respect the function with the property $f(q^j)=f(q^{-j})$ at arbitrary $j$. However, a non-trivial function with such a property can not be a polynomial.

One definitely can require that it is periodic, $f(q^j)=f(q^{-j})$ at any $q=\exp\left({2\pi ik\over n}\right)$ with $n\le 2m$. However, this is far not enough for fixing the form (\ref{ellcf}).

\subsection{Rodrigues type formula}

\paragraph{Non-elliptic case.} In the non-elliptic case, the BA function can be obtained via a counterpart of the Rodrigues formula \cite{Cha}:
\be\label{prod}
\Psi_m(z_k,\vec\lambda)\sim\prod_{\substack{k_{ij}=0 \\ \sum_{i<j}k_{ij}^2\ne 0}}^m\left(\hat D_\Z-\underbrace{q^{-{m\over 2}}\sum q^{\lambda_i+\sum_{j>i}k_{ij}
-\sum_{j<i}k_{ji}}}_{\Lambda(\lambda_i+\sum_{j>i}k_{ij}
-\sum_{j<i}k_{ji})}\right)
\left[q^{(\vec\lambda+m\vec\rho)\cdot\vec z}\Delta^{-1}_{q,t}(\Z)\right]
\ee
where the product runs over all \textit{non-zero} multiplicity vectors for positive roots
$k_{ij}$, $i<j, i,j=1..N$.

Here
\be
\Delta_{q,t}^{-1}(\Z):=\prod_{i\ne j}\prod_{k=1}^m\left(q^k{\Z_i\over \Z_j}-1\right)
\ee
is (proportional to) the inverse of the $q,t$-deformed square of the Vandermonde determinant at $t=q^{-m}$. The BA function is an eigenfunction of the Ruijsenaars-Macdonald operator with the eigenvalue $\Lambda(\vec\lambda)$:
\be
\hat D_\Z\Psi_m(z_k,\vec\lambda)=\Lambda(\vec\lambda)\cdot\Psi_m(z_k,\vec\lambda)
\ee
In particular, in the $N=2$ case, the formula acquires the form
\be\label{prodN2}
\Psi_m(\vec z,\vec\lambda)\sim\prod_{j=1}^m\left(\hat D_{\Z}-q^{-{m\over 2}}\sum q^{\lambda_i+2j\rho_i}\right)
\left[q^{(\vec\lambda+m\vec\rho)\cdot\vec z}\Delta^{-1}_{q,t}(\Z)\right]
\ee
The idea of the proof of formula (\ref{prod}) is that
\be\label{Q}
q^{(\vec\lambda+m\vec\rho)\cdot\vec z}\Delta^{-1}_{q,t}(\Z)\sim q^{(\vec\lambda+m\vec\rho)}\cdot
Pol(x)
\ee
is a (quasi)polynomial of $q^{z_i}$ with the periodicity property (\ref{symm1}). At the same time, the polynomial $Pol(x)$ contains too many terms: it has the structure as in (\ref{gsBA}) with $k_{ij}$ running from $-m$ to $m$. However, action of $\hat D_{\Z}-\alpha$ with constant $\alpha$ preserves the periodicity property, and one can choose a particular constant $\alpha$ in such a way that the number of terms with negative $k_{in}$ in $Pol(x)$ is reduced. Repeating this procedure $m$ times, one reduces the polynomial in (\ref{Q}) to a new polynomial $\widehat{Pol}(x)$ of the form (\ref{gsBA}) with the proper ranges of summation so that the obtained expression still celebrates the periodicity property (\ref{symm1}), i.e. it is the BA function (up to a normalization factor).

\paragraph{Elliptic case.} Similar arguments are clearly not applicable in the case of elliptic deformation.
Nevertheless, one could hope that there is a formula of the form
\be\label{prodN2ell}
\Psi_m(\vec z,\vec\lambda;q,w)=\widehat{\mathfrak{D}}
\left[q^{(\vec\lambda+m\vec\rho)\cdot\vec z}\Delta^{-1}_{q,t,w}(\Z)\right]
\ee
with a properly defined $\Delta^{-1}_{q,t,w}(\Z)$ and an operator $\widehat{\mathfrak{D}}$. The main problem is that the elliptic extension of the inverse square of $q,t$-Vandermonde determinant \cite{MMell} (see also \cite{NZ} at $t=q$),
\be\label{ellV}
\Delta_{q,t,w}^{-1}(\Z)\stackrel{?}{\sim}
\prod_{i\ne j}{\Gamma(\Z_i/\Z_j;q,w)\over \Gamma(t\Z_i/\Z_j;q,w)}\stackrel{t=q^{-m}}{=}
\prod_{j=1}^m\theta_w(q^{-j}\Z_i/\Z_j)\sim \prod_{j=1}^m\theta_w(q^{j}\Z_i/\Z_j)
\ee
is not a polynomial but an infinite series in $\Z_i$, and it is not simple to construct an operator $\widehat{\mathfrak{D}}$ that, upon acting on (\ref{ellV}), gives rise to the (quasi)polynomial elliptic BA function.

\section{Orthogonality conditions}

Let us explain in other terms why one should not expect simple linear equations in the elliptic case.
To this end, we first discuss that non-elliptic Macdonald polynomials are orthogonal w.r.t. to two scalar products scalar product, while the elliptic Macdonald polynomials are not. Then, we explain that the non-elliptic BA functions are also orthogonal w.r.t. to one of these scalar products (the other one is formulated in terms of power sums, i.e. is applicable to symmetric polynomials only), while the elliptic BA functions do not. In the meanwhile, the orthogonality conditions are related to the linear conditions.

\subsection{Orthogonality of elliptic Macdonald polynomials}

\paragraph{Scalar product in terms of power sums.} The non-elliptic Macdonald polynomials satisfy the orthogonality relations
\be\label{or}
\Big<M_\lambda,M_\mu\Big>\ne 0\ \ \ \ \ \ \hbox{iff }\lambda=\mu
\ee
where the scalar product is defined as follows. Being symmetric polynomials, the non-elliptic Macdonald polynomials can be reformulated in terms of power sums $p_k=\sum_{i=1}^Nx_i^k$., and the scalar product in these terms is given by \cite{Mac}:
\be\label{Ssp1}
\Big<p_\Delta,p_{\Delta'}\Big>_{M_1}:=\delta_{\Delta,\Delta'}z_\Delta\prod_{i=1}^{l_\Delta}{q^{\delta_i}-1\over t^{\delta_i}-1}
\ee
where $z_\Delta:=\prod_k k^{m_k}m_k!$ is the standard symmetric factor of the Young diagram (order of the automorphism). This scalar product is extended to the whole polynomials by linearity. It can be also given for arbitrary polynomials $f(\{p_k\})$ and $g(\{p_k\})$ of $p_k$'s by the manifest action
\be
\Big<f,g\Big>_{M_1}=
\left.f\left(\left\{k{q^k-1\over t^k-1}{\p\over\p p_k}\right\}\right)g(\{p_k\})\right|_{p_k=0\ \forall k}
\ee

\paragraph{Bi-orthogonal system of elliptic Macdonald polynomials.} The orthogonality relation (\ref{or}) is, however, not extendable to the elliptic case \cite{AKMM3}: in this case one can only construct a system of polynomials orthogonal to the elliptic Macdonald polynomials $P_\mu(\{p_k\};q,t,w)$. It makes sense to construct such a bi-orthogonal system with slightly different scalar product\footnote{It can be also given by the manifest action
$$
\Big<f,g\Big>_{S_1}=\left.f\left(\left\{k{\p\over\p p_k}\right\}\right)g(\{p_k\})\right|_{p_k=0\ \forall k}
$$
}
\be\label{Ssp2}
\Big<p_\Delta,p_{\Delta'}\Big>_{S_1}:=\delta_{\Delta,\Delta'}z_\Delta
\ee
which gives rise to orthogonality of the Schur functions. Denote the polynomials orthogonal to $P_\mu(\{p_k\};q,t,w)$ w.r.t. the scalar product (\ref{Ssp1}) through $P^{\perp}_\mu(\{p_k\};q,t,w)$:
\be\boxed{
\Big<P_\lambda,P^\perp_\mu\Big>_{S_1}=\delta_{\lambda\mu}}
\ee
For the sake of convenience and following \cite{AKMM3}, we introduce relabeled polynomials
\be\boxed{
E_\mu(\{p_k\};q,t,w):=P^{\perp}_{\mu^\vee}(\{(-1)^{k+1}p_k\};t,q,w)}
\ee
where $\mu^\vee$ denotes the conjugate partition. These polynomials have a series of distinguished properties \cite{AKMM3,MMZ}:
\begin{itemize}
\item
First of all, their composition rules coincide with those of the Schur functions and non-elliptic Macdonald polynomials, i.e. the generalized Littlewood-Richardson coefficients (coefficients of expansion of the product of two different polynomials $P^{\perp}_\mu(\{p_k\};q,t,w)$ into a sum of $P^{\perp}_\mu(\{x_i\};q,t,w)$) are vanishing iff vanishing are the corresponding Littlewood-Richardson coefficients for the Schur functions (i.e. they respect the standard structure of products of $SL_N$ representations) \cite{AKMM3}.
\item Second, these polynomials are eigenfunctions of the Hamiltonians dual to the elliptic Ruijsenaars Hamiltonians \cite{MMZ,MMdell}. Note that, in the limit $w\to 1$, $E_\mu(\{p_k\};q,t,w)=M_\mu(\{p_k\};q,t)$ are the non-elliptic Macdonald polynomials, while, at $t=q$, $E_\mu(\{p_k\};q,q,w)=S_\mu(\{p_k\})$ are the Schur functions.
\item Algebraically, the polynomials $E_\mu(\{p_k\};q,t,w)$ realize a basis in the vertical Fock representation of the elliptic DIM algebra \cite{MMZ}.
\end{itemize}

\subsection{Another scalar product}

Our previous consideration was completely in terms of power sums and, hence, is applicable to the symmetric polynomials only. Now we note that there is another scalar product \cite[sec.VI.9]{Mac}, which is suitable for any polynomials, including the BA functions. It is defined for two functions $f(\{x_i\})$ and $g(\{x_i\})$ of $N$ variables $x_i$ as
\be\label{asp1}
\Big<f,g\Big>_{M_2}:={1\over N!}\left(\prod_{i=1}^N\oint{dz_i\over z_i}\right)
f(x)g(x^{-1})\prod_{i\ne j}{\Big({x_i\over x_j};q\Big)_\infty\over\Big(t{x_i\over x_j};q\Big)_\infty}
\stackrel{t=q^k}{=}{1\over N!}\left(\prod_{i=1}^N\oint{dz_i\over z_i}\right)
f(x)g(x^{-1})\prod_{i\ne j}\prod_{r=0}^{k-1}\left(1-q^r{x_i\over x_j}\right)\nn\\
\ee
With this product,
\be\label{M2sp}
\Big<M_\lambda,M_\mu\Big>_{M_2}=||M_\lambda||^2\delta_{\lambda\mu}
\ee
where
\be
||M_\lambda||^2&=&\prod_{i<j}\prod_{r=0}^\infty{(q^{\lambda_i-\lambda_j+r}t^{j-i}-1)\over(q^{\lambda_i-\lambda_j+r}t^{j-i+1}-1)}
{(q^{\lambda_i-\lambda_j+r+1}t^{j-i}-1)\over(q^{\lambda_i-\lambda_j+r+1}t^{j-i-1}-1)}\stackrel{t=q^k}{=}
\prod_{r=0}^{k-1}{(q^{\lambda_i-\lambda_j+r+k(j-i)}-1)\over(q^{\lambda_i-\lambda_j-r+k(j-i)}-1)}
\ee
At $t=q$, this scalar product becomes the Schur scalar product:
\be\label{asp2}
\Big<S_\lambda,S_\mu\Big>_{S_2}:={1\over N!}\left(\prod_{i=1}^N\oint{dz_i\over z_i}\right)
S_\lambda(x)S_\mu(x^{-1})\prod_{i\ne j}\left(1-{x_i\over x_j}\right)=\delta_{\lambda\mu}
\ee
Now, what one can see is that the pair of elliptic polynomials $P_\mu$ and $E_\nu$ is a bi-orthogonal w.r.t. to this Schur scalar product as well:
\be
\Big<S_\lambda,S_\mu\Big>_{S_2}=\delta_{\lambda\mu}
\ee
Note that $P_\mu^\perp$ is orthogonal to $P_\nu$ in (\ref{P2}) in the scalar products (\ref{asp2}),
\be\boxed{
\Big<P_\lambda,P^\perp_\mu\Big>_{S_2}=\delta_{\lambda\mu}}
\ee
{\bf only if} one restricts $\mu$ by the partitions with no more than $N$ parts. The reason is that $P_\mu^\perp$ does not automatically vanish if $\mu$ has more than $N$ parts, in contrast with $P_\mu$, $M_\mu$ and $S_\mu$. At the same time, $E_\mu$ does vanish if $\mu$ has more than $N$ parts, but it is not orthogonal to $P_\mu$ with the scalar product (\ref{asp2}).

\paragraph{An example of $N=2$.} In fact, one can find the general expression for $E_\mu$ at $N=2$.
However, as we explained, the orthogonality conditions with the scalar product (\ref{asp2}) are not enough to fix $P_\mu^\perp$ unambiguously. Instead, one can evaluate $P_\mu^\perp$ using the scalar product in terms of power sums (\ref{Ssp2}) and then reduce the polynomial to the case of two variables $p_k=x_1^k+x_2^k$, the results for $E_\mu$ reads
\be
E_{[n]}=\sum_{k=0}^nx_1^{n-k}x_2^k\sum_{i=0}^k\binom{n-2i}{i}\sum_{0\le a_1\le n-2}
\sum_{0\le a_2\le a_1-2}\ldots\sum_{0\le a_i\le a_{i-1}-2}\prod_{j=1}^i{\theta_w(q^{a_i+1})
\theta_w(t^2q^{a_i})\over\theta_w(tq^{a_i+1})\theta_w(tq^{a_i})}
\ee
where $\binom{n-2i}{i}$ is the binomial coefficient.

However, it is important to note that, even in this example, the $E$-polynomials do not reduce to $2m$ terms at $t=q^{-m}$. Hence, it is not simple to extract a part that can be continued to non-integer $n$, and the notion of BA function in this case is not clear.

\subsection{Obtaining the non-elliptic BA function from the orthogonality condition}

One can construct the non-elliptic Macdonald polynomials from the orthogonality condition. To this end, one represents them by triangle expansion in monomial symmetric polynomials $m_\mu$,
\be
M_\lambda=\sum_{\mu\le\lambda}K_\lambda^\mu(q,t)m_\mu
\ee
where $K_\lambda^\mu(q,t)$ are Kostka coefficients, the partitions are partially ordered in accordance with the dominance rule, and $K_\lambda^\lambda(q,t)=1$. The Kostka coefficients then unambiguously determined from the orthogonality condition (\ref{M2sp}).

Now notice that
\be
m_\mu=\sum_{w\in W}wx^{\lambda}
\ee
where the sum goes over all elements $w$ of the Weyl group of $A_{N-1}$, and $x^\lambda:=\prod_{i=1}^Nx_i^{\lambda_i}$. On the other hand, the non-elliptic Macdonald polynomial is related up to a normalization factor to the non-elliptic BA function via the formula
\be
M_{\vec\lambda+m\vec\rho}(\{x_i\};q,q^{-m})=\sum_{w\in W}\Psi_m(wx,\lambda)
\ee
Hence, one obtains that
\be\label{psi}
\Psi_m(x,\lambda-m\vec\rho)=\sum_{\mu\le\lambda}K_\lambda^\mu(q,t)x^\mu
\ee
In the meanwhile, the definition of the BA function,
\be\label{gsBA2}
\Psi_m(\vec x,\vec\lambda)=x^{\vec\lambda+m\vec\rho}\ \sum_{k_{ij}=0}^m\prod_{i>j}\left({x_i\over x_j}\right)^{k_{ij}}\psi_{m,\vec\lambda,k}
\ee
means that the sum of all $\psi$'s corresponding to one and the same $x^\mu$ is equal to the Kostka coefficient, and, hence, the linear equations defining them are somehow equivalent to the orthogonality condition. Note that the scalar product (\ref{asp1}), at $t=q^{-m}$, has the form
\be
\Big<f,g\Big>_{M_2}={1\over N!}\left(\prod_{i=1}^N\oint{dz_i\over z_i}\right)
{f(x)g(x^{-1})\over\prod_{i\ne j}\prod_{k=1}^m\Big(1-q^k{x_i\over x_j}\Big)}
\ee
and, when $\lambda$ and $\mu$ are partitions (in fact, their difference is enough to be a partition), the BA functions are also orthogonal w.r.t. to this scalar product \cite[sec.4]{ChE}:
\be
  \oint \prod_{i=1}^N{dx_i\over x_i}{\Psi_m(x,\lambda)\Psi_m(x^{-1},\mu)\over\prod_{i\ne j}\prod_{k=1}^m\Big(1-q^k{x_i\over x_j}\Big)}=||\Psi_m(\lambda)||^2\cdot\delta_{\lambda\mu}
\ee
where
\be
||\overline{\Psi}_m(\lambda)||^2&=&q^{-m\lambda-m(m+1)}\cdot\prod_{j=1}^m(q^{\lambda+j}-1)(q^{\lambda-j}-1)\nn\\
||\Psi_m(\lambda)||^2&=&(-1)^mq^{-m(m+1)\over 2}\cdot\prod_{i\ne j}\prod_{r=1}^m(q^{\lambda_i-\lambda_j+r}-1)
\ee
In other words, one has the matrix (where we define the integration first over $x_N$, then over $x_{N-1}$, etc)
\be
\Lambda_{\mu,\mu'}^{(N,m)}:=\Big<x^\mu,x^{-\mu'}\Big>=\oint \prod_{i=1}^N{dx_i\over x_i}{\prod_{i=1}^N x_i^{\mu_i-\mu_i'}\over\prod_{i\ne j}\prod_{k=1}^m\Big(1-q^k{x_i\over x_j}\Big)}\sim
\delta_{_{\sum_i (\mu_i-\mu'_i),0}}
\ee
and the triangle matrix of the Kostka coefficients diagonalizes it.

The problem of finding the Kostka coefficients via orthogonality conditions in order to construct the BA function is equivalent to linear equations. Unfortunately, as we explained in this section, the orthogonality {\bf conditions} are not available for the BA functions in the elliptic case, as are the linear equations. Indeed, what one can do is to define a {\bf new} function orthogonal to the elliptic BA function similarly to constructing the {\bf new }polynomials $P^\perp_\mu$ from $P_\mu$. However, it does not somehow restrict the polynomials $P_\mu$, neither restricts the elliptic BA functions.

\section{Conclusion}

In this letter, we presented an evidence that the entire Macdonald triad possesses an elliptic deformation.
Known before were one of the generalizations \cite{KS} of the Ruijsenaars Hamiltonians, the
ELS functions \cite{AKMM2} being their eigenfunctions \cite{MMZ},
and the reduction of these latter to two orthogonal sets of elliptic polynomials \cite{AKMM2,MMZ},
even more mysterious than the usual Macdonald polynomials.
In fact, the system of commuting Hamiltonians is fully known only for one of the sets,
and those for the other one still remain to be found \cite{MMdell}.

The main claim of this letter is that the ``Baker Akhiezer functions'',
the technically simplest ingredient of the triad can also be elliptically deformed.
Natural, however, is only the deformation of {\it the explicit expression}:
it remains a nice polynomial reduction of elliptic Macdonald and Noumi-Shiraishi functions.
As to the conditions defining the Baker-Akhiezer function independently on these functions,
i.e. to the system of equations replacing the linear equations in the non-elliptic case,
it is not unique, and the question remains about the ways to fix this ambiguity.
This question looks promising, because it provides a chance to understand
the true meaning of these conditions,
which appeared just from nowhere in \cite{Cha}, were demonstrated to be very powerful,
possess spectacular generalizations to ``twisted'' BA functions \cite{ChE,ChF}, and match \cite{MMP,MMP1,MMP2} representation theory of the DIM algebra \cite{DIM}.
A problem of twisted elliptic BA function and their relation to Abelian subalgebras of the elliptic DIM algebra \cite{Saito,ell},
i.e. to a possible set of compatible integrable elliptic systems is now also in agenda.
We hope that all these questions will now attract attention and will be solved in foreseeable future.

\section*{Acknowledgements}

This work was supported by the Russian Science Foundation (Grant No.23-41-00049).

\end{document}